\documentclass[12pt]{iopart}

\usepackage{iopams}

\newcommand{\bra}[1]{\left\langle #1\right|}
\newcommand{\ket}[1]{\left| #1\right\rangle}
\newcommand{\braket}[1]{\langle #1\rangle}
\newcommand{\g}{{\mathbf 0}}
\newcommand{\x}{{\mathbf 1}}
\newcommand{\ii}{{\mathbf i}}

\usepackage[usenames]{color}

\newcommand{\pr}[1] {\ket{#1}\bra{#1}}
 
\begin{document}
\title{Bound entanglement in the Jaynes-Cummings model}
\author{Nicol\'as Quesada}
\address{McLennan Physical Laboratories, Institute for Optical Sciences and Centre for Quantum Information and Quantum Control, University of Toronto, 60 St. George Street, Toronto, Ontario, Canada M5S 1A7}
\ead{nquesada@physics.utoronto.ca}

\author{Anna Sanpera}
\address{ICREA, Instituci\`o Catalana de Recerca i Estudis Avan\c{c}ats, E-08010 Barcelona, Spain.}
\address{Departament de F\'{i}sica, Universitat Aut\`{o}noma de Barcelona, E-08193 Bellaterra, Spain.}

\date{\today}

\begin{abstract}
We study in detail entanglement properties of the Jaynes-Cummings model assuming a two-level atom (qubit) interacting with the first  $N$ levels of an electromagnetic field mode (qudit) in a cavity. In the Jaynes-Cummings model, the number operator is the conserved quantity that allows for the exact diagonalization of the Hamiltonian and thus we study states that commute with this conserved quantity and whose structure is preserved under the Jaynes-Cummings dynamics. Contrary to the common belief, we show that there are bound entangled states that satisfy the symmetries imposed by the conservation of the number of excitations when $N>3$. Furthermore we show that \emph{the Jaynes-Cummings interaction can be used to generate bound-entanglement} between the atom and the mode. 
\end{abstract}

\maketitle

\section{Introduction}
The Jaynes-Cummings (JC) model is one of the workhorses of quantum optics research\cite{Jaynes63,Knight93}. In this model the interaction of a two level atom with a single electromagnetic field (EMF) mode is studied under the dipole and rotating wave approximations. Its rich dynamics has been used to model and understand phenomenology both in the realm of cavity\cite{Rempe87,Brune96,Haroche01} and circuit\cite{Blais04} QED and in trapped ions\cite{Wineland96,Onofrio97}, where the EMF is replaced by the quantized vibrations of the ions. There is also a significant amount of literature devoted to entanglement in the JC model. While some of the previous results concerned pure states\cite{Knight92,Liu07,Zhang10}, mixed state entanglement has been addressed using either entropic relations\cite{Tannor05,Guo11,Cummings99,Lendi98,Obada04}  or detection techniques drawn directly from the field of quantum information. In particular, projecting the Fock space of the EMF mode onto a two dimensional subspace results into  a combined Hilbert space that essentially reduces to a two-qubit system whose entanglement properties are well understood\cite{Rendell03,Cai05,Kayhan11,Masanori99,Bose01,Vera09}. Also, the partial transpose criterion (PPT) introduced by Peres \cite{Peres96} has been used to this aim ~\cite{Struntz05,Plenio03,Farsi07,Quesada11,Suarez12}. We note, however, that neither the use entropic inequalities (which are a corollary of the majorization criterion\cite{Nielsen01}) or the projection of the system onto  $2\times2$ subspaces leads to entanglement criteria that are stronger than the PPT one\cite{Hiroshima03,Chen12}. Thus, so far only ``distillable'' (or free) entanglement has been detected in these systems \cite{Horodeckireview}.\\
In this paper we first formalize the problem that most previous studies have analyzed in terms of the conservation of the number of excitations that makes the JC Hamiltonian an exactly solvable problem. We then show that this symmetry gives rise to a super selection rule\cite{Schuch04} that severely constrains the structure of the density matrices describing the states.
With these tools at hand we review entanglement detection techniques previously used and show that they are, in general, weaker than the Peres criterion. Further, we show numerically that other entanglement criteria that are not based on the use of positive maps but not completely positive ones are not able to supplement the Peres criterion. This is the case of the computable cross-norm or realignment (CCNR) criterion\cite{Rudolph03,Guhne09} and some covariance matrices (CM) corollaries\cite{Zhang08,Oleg08}. We construct explicitly states that are PPT but entangled demonstrating that there exist bound entangled states in the JC model that are not detected by any of the previous methods and that some of these states can be generated by the JC interaction. We emphasize that to the best of our knowledge, this is the first study that addresses the possibility of generating bound entanglement using the JC dynamics which is a commonly occurring interaction in different physical systems. We note however that several theoretical\cite{geza07,patane07,ferraro08} and experimental\cite{Amselem09,lavoie10,barreiro10,di11} investigations have been performed seeking bound entanglement in other physical systems.

This paper is organized as follows: In section \ref{sec:moti}, we introduce and define the class of states to be studied here and argue that those are the natural mixed states resulting from the dynamics of the JC model. Further, we introduce the notation that will be used in the rest of the paper in terms of qubit-qudit ($2\times N$) density operators. In section \ref{sec:dete}, we derive under which conditions these states are PPT, \emph{i.e.} have a positive partial transpose. 
In section \ref{sec:bound}, we focus at $N =4$ and show that despite the high symmetry of the states considered bound entangled states exist.  Following some ideas presented in \cite{kraus00,tura12}, we demonstrated the existence of bound entanglement by using the range criterion \cite{Horo97}.
Therefore, under the JC dynamics, the PPT criterion is necessary but not sufficient to ensure entanglement. In section \ref{sec:dyn} we show how using the JC interaction it is possible to generate bound entangled states starting from uncorrelated ones. Final remarks are given in section \ref{sec:conc} . 
For completeness, in \ref{app:crit} we present two well known entanglement criteria that have been used to detect bound entanglement and numerically show that they are ineffective for the states considered here and in \ref{app:hull} we construct the convex hull for the PPT separable states when $N=2,3$.

\section{Motivation and Definitions}\label{sec:moti}
Assume a cavity with a finite quality factor where a $N-1$ photon state, $\ket{N-1}$, has been prepared. After such initial preparation an atom in its ground state $\ket{\g}$ enters the cavity and starts to interact with the photons. Typically, three processes will occur in the dynamics of the system:
\begin{itemize}
\item The atom and the cavity will reversibly exchange one excitation which under the assumptions of the JC model will causes the transition $\ket{\g}\otimes\ket{N-1} \Leftrightarrow \ket{\x}\otimes \ket{N-2}$ with $\ket{\x}$ being the excited state of the atom.
\item The cavity might irreversibly lose one of its photons since it has a finite quality factor.
\item Once the atom is excited it can spontaneously and irreversibly emit a photon that escapes the cavity.
\end{itemize}
As time goes by, the dynamics of the system is in general quite complicated and to treat it completely it is necessary to include the modes outside the cavity that gives rise to an irreversible behavior. Nevertheless, under very general assumptions such as that the interaction with the outside modes preserves the number of photons (\emph{i.e.} that each photon that disappears from the cavity becomes a photon in one of the outside modes) the state of the system commutes with the number of excitations of the atom plus cavity. Including the first $N$ Fock states in the dynamics one can write the number of excitations operator as:
\begin{eqnarray}\label{def}
\Pi&=&\Pi_2 \otimes \mathbb{I}+ \mathbb{I} \otimes \Pi_N,\\
\Pi_2&=&\sum_{\ii = \g}^\x \ii \ket{\ii}\bra{\ii}=\ket{\x}\bra{\x},\nonumber \quad 
\Pi_N=\sum_{n=0}^{N-1}n\ket{n}\bra{n}.\nonumber
\end{eqnarray}
At any time the state of the atom and photons $\rho$ satisfies:
\begin{eqnarray}\label{comm}
\left[\rho, \Pi \right]=0.
\end{eqnarray}
The above symmetry gives rise directly to the following super-selection rule\cite{Schuch04}:
\begin{eqnarray}\label{super}
\bra{\ii  n} \rho \ket{\mathbf{j} m} \propto \delta_{\ii+n,\mathbf{j}+m},
\end{eqnarray}
where as usual $\ket{\ii n} \equiv \ket{\ii}\otimes\ket{n}$ and $\ket{\ii},\ket{\mathbf{j}}$ are two atomic states and $\ket{m},\ket{n}$ are Fock states with $m$ and $n$ photons respectively.
It can be easily shown that if at $t=0$ the state $\rho(t=0)$ satisfies (\ref{comm}) and if it evolves under the von Neumann equation with the JC Hamiltonian,
\begin{eqnarray}\label{neu}
\frac{d}{dt} \rho=i\left[\rho, H_{JC}\right],
\end{eqnarray}
then at later times it also satisfies (\ref{comm}). In (\ref{neu}) the JC Hamiltonian is given by:
\begin{eqnarray}\label{HJC}
H_{JC}&=&\omega_0 \mathbb{I} \otimes a^\dagger a+\left(\omega_0-\Delta\right) \sigma^\dagger \sigma \otimes \mathbb{I}-i g\left(\sigma \otimes a^\dagger-\sigma^\dagger \otimes a \right),
\end{eqnarray}
where $a, (a^\dagger)$ is a  annihilation  (creation) bosonic operator satisfying $[a,a^\dagger ]=\mathbb{I}$,  $\sigma=\ket{\g}\bra{\x}$, $g$ is the light-matter coupling constant (the Rabi frequency), $\omega_0$ is the frequency of the mode and $\Delta$ is the detuning between the mode and the transition frequency between the two atomic states.
We note that the JC Hamiltonian also commutes with the number operator when all the states in the Fock ladder are included (\emph{i.e.} $N \to \infty$ in (\ref{def})) and thus it is reasonable to think of the states $\rho$ in (\ref{comm}) as the natural states of the JC dynamics. Even more interesting is the fact that when some phenomenological dissipation terms are added to the RHS of (\ref{neu}) the dynamics still preserves the structure of the density matrix describing the state. This is the case if Lindblad terms of the form:
\begin{eqnarray}
\mathcal{L}_{\sqrt{\gamma_O}O}\{\rho\}=\gamma_O\left(O \rho O^\dagger -\frac{\rho O^\dagger O+O^\dagger O \rho}{2} \right),
\end{eqnarray}
with $O= \mathbb{I} \otimes a$ for photon depopulation, $O=\mathbb{I} \otimes a^\dagger$ for photon re-population, $O=\mathbb{I} \otimes a^\dagger a$ for photon dephasing, $O=\sigma \otimes \mathbb{I}$ for atom depopulation, $O=\sigma^\dagger  \otimes \mathbb{I} $ for atom re-population, $O=\sigma^\dagger \sigma  \otimes \mathbb{I} $ for atomic dephasing are added. 
Even if a microscopic treatment is used and a rigorous derivation of the system operator dynamics is performed by first coupling it to a bosonic reservoir and then tracing it out, if the dynamics of the system plus reservoir as a whole preserves the overall number of excitations the shape of the states satisfying (\ref{comm}) will be preserved \cite{Scala07}.\\

For future convenience we name the non-zero elements of the density matrix that satisfy the super-selection rule (\ref{super}) as follows. The two sets of $N$ nonzero elements along the diagonal (populations) we label:
\begin{eqnarray}\label{not1}
a_n \equiv \bra{\g n }\rho \ket{\g n} \quad b_n \equiv \bra{\x n}\rho \ket{\x n},
\end{eqnarray}
and the two sets of $N-1$ non-zero off-diagonal elements or coherences we parametrize as:
\begin{eqnarray}\label{not2}
c_n \equiv \bra{\g n} \rho \ket{\x n-1} \quad c_n^\star  \equiv \bra{\x n-1}\rho \ket{\g n}.
\end{eqnarray}
In the above equation it is assumed that $\rho$ is a state and thus it is automatically Hermitian. The state is explicitly given by:
\begin{eqnarray}\label{state}
\rho&=&\sum_{n=0}^{N-1}  \left(a_n\ket{\g n}\bra{\g n}+b_n\ket{\x n}\bra{\x n}\right) \\
&&+ \sum_{n=1}^{N-1} \left(c_n^\star  \ket{\x n-1}\bra{\g n}+c_n\ket{\g n}\bra{\x n-1}\right). \nonumber
\end{eqnarray}
It can be represented as a $2\times 2$ matrix in the atom-field basis whose entries are $N\times N$ sparse matrices:
\begin{eqnarray}
\rho=\left(
\begin{array}{c|c}
 A & C \\
\hline
 C^{\dagger } & B \\
\end{array}
\right),
\end{eqnarray}
more explicitly in the basis given by the tensor product of $\left\{\ket{\ii} \right\}_{\ii=\g}^{\x}$ and $\left\{\ket{n} \right\}_{n=0}^{N-1}$ the matrix has the following structure
\begin{eqnarray}\label{bigmat}
\rho = \left(
\begin{array}{cccc|cccc}
 a_0 &  &  &  &  &  &  &  \\
  & \ddots &  &  & c_1 &  &  &  \\
  &  & \ddots &  &  & \ddots &  &  \\
  &  &  & a_{N-1} &  &  & c_{N-1} &  \\
\hline
  & c_1^\star  &  &  & b_0 &  &  &  \\
  &  & \ddots &  &  & \ddots &  &  \\
  &  &  & c_{N-1}^\star  &  &  & \ddots &  \\
  &  &  &  &  &  &  & b_{N-1}\\
\end{array}
\right).
\end{eqnarray}
In spite of its simplicity,  the separability properties of such state are intricate. Notice that although the above matrix was constructed to be explicitly hermitian, to represent a physical state the density operator must also be positive semi-definite, which is equivalent to require that: 
\begin{eqnarray}\label{pos}
 a_n,b_n \geq 0, \quad  |c_n|^2 \leq a_n b_{n-1}. 
\end{eqnarray}
The positivity conditions (\ref{pos}) guarantees that all the eigenvalues of $\rho$ will be non-negative. 
Moreover, unless at least one of the inequalities  (\ref{pos}) is saturated $\rho$ is of full rank.
Notice that by applying a local unitary transformation to the density matrix it is possible to make all the coefficients in it non-negative. Such local unitary is given by
\begin{eqnarray}\label{local}
 \mathcal{V}=\sum_{n=0}^{N-1} e^{i \theta_n} \ket{n}\bra{n}, \quad \theta_n-\theta_{n-1}=\arg(c_n),
\end{eqnarray}
and since the entanglement properties of the state are invariant under local unitaries it suffices to consider only the absolute values of the $c_n$. In other words, from now on we deal with density operators whose matrix elements are non-negative.

\section{Entanglement detection}\label{sec:dete}
To determine if the state (\ref{bigmat}) is entangled or not \emph{i.e.} whether it can be written as:
\begin{eqnarray}
\rho=\sum_i p_i \ket{\phi_i}\bra{\phi_i} \otimes \ket{\xi_i}\bra{\xi_i},
\end{eqnarray}
is not a trivial question since the state acts in a Hilbert space of dimension $2N$ where only sufficient but not necessary separability criteria are known. 
Furthermore, one would like to quantify the extent to which the state cannot be written as a separable mixture. One often used metric is the concurrence $\mathcal{C}\left(\rho \right)$~\cite{Wootters98}. For a pure state $\psi=\ket{\psi}\bra{\psi}$ in a $2 \times N$ system the concurrence is simply:
\begin{eqnarray}
\mathcal{C}(\psi )=\sqrt{2\left(1-\tr\left(\mu^2\right)\right)}\\
\mu=\tr_{A/B}(\psi),
\end{eqnarray}
where $\tr_{A/B}$ denotes the partial trace over subsystem $A$ or $B$. For mixed states the concurrence is extended via the convex roof construction:
\begin{eqnarray}
\mathcal{C}(\rho)=\min_{\sum_i p_i \psi_i =\rho} \sum_i p_i \mathcal{C}(\psi_i),
\end{eqnarray}
with $p_i$ being probabilities and the $\psi_i$ being pure states. This quantity ranges from zero (for separable states) to one for maximally entangled ones.\\
A possible approach to answer whether a state is separable or not is to obtain the partial transpose (PT) with respect one of its subsystems (does not matter which one) and see if it has negative eigenvalues\cite{Peres96}: if there is at least one negative eigenvalue then $\rho$ must be entangled. 
The PT of $\rho$ with respect to subsystem $B$ reads: 
\begin{eqnarray}
\rho^\Gamma=\left(
\begin{array}{c|c}
 A & C^T \\
\hline
 C^{\star} & B \\
\end{array}
\right).
\end{eqnarray}
We point out that once the local unitary (\ref{local}) has been applied to $\rho$ then $\rho^{T_A}=\rho^{T_B}\equiv \rho^\Gamma$, where $\rho^{T_i}$ is the partial transpose with respect to the $i^{th}$ party.
For $\rho^\Gamma$ to be positive semi-definite one new constraint is needed:
\begin{eqnarray}\label{postrans}
|c_n|^2 \leq a_{n-1} b_n.
\end{eqnarray}
The violation of the constraints above gives a sufficient condition for $\rho$ to be entangled. One reaches the same condition by finding an upper bound to the concurrence of the state projecting the qudit into the subspace spanned by two consecutive Fock states $\ket{n},\ket{n+1}$ \cite{Rendell03}. 
Another way to arrive to the constraints given by (\ref{postrans}) is by using the bound on the concurrence for a $2\times N$ dimensional system derived in \cite{gerjuoy03} in which a set of $N(N-1)/2$ quantities are proposed to bound the concurrence of the system. Of these only $N-1$ turn out to be not trivial and they express the same bounds contained in (\ref{postrans}). Using the results from \cite{gerjuoy03} the following bound for the concurrence is found:
\begin{eqnarray}\label{ger}
\mathcal{C}(\rho)\geq \mathcal{G}(\rho)=\sqrt{\sum_{l=1}^{N-1} \mathcal{G}_l(\rho)^2} \\
\mathcal{G}_l(\rho)=2\max\left\{0,|c_l|-\sqrt{a_{l-1}b_l} \right\} 
\end{eqnarray}
It was pointed out in \cite{Chen12} that the above bound cannot have more entanglement detection capabilities than the positive PT criterion, it is interesting to notice that for these states they have \emph{exactly} the same entanglement detection capabilities.
The Negativity\cite{Vidal02} can also be used to quantify the failure of $\rho^\Gamma$ to be positive semi-definite:
\begin{eqnarray}\label{neg}
\mathcal{N}(\rho)&=& -\sum_{n=1}^N \min \left\{0, \lambda^\Gamma_{n-} \right\},\\
\lambda^\Gamma_{n-}&=&a_{n-1}+b_n-\sqrt{(a_{n-1}-b_n)^2+4|c_n|^2}.
\end{eqnarray}
The normalization is such that it takes a value of $1$ when $\rho$ is a maximally entangled pure state $\rho=\ket{\psi}\bra{\psi}$ with $\ket{\psi}=\frac{1}{\sqrt{2}}\left( \ket{\g n}+\ket{\x n-1} \right)$. It was pointed out in \cite{Kai05} that the negativity with the normalization used above is another lower bound for the concurrence:
\begin{eqnarray}
\mathcal{C}(\rho) \geq \mathcal{N}(\rho).
\end{eqnarray}
When the Negativity (or for these states the bound (\ref{ger})) is zero and $N>3$ the criterion is inconclusive \cite{Horo96}, \emph{i.e.} it is known that in the general case of  $2\times N$ there are entangled states not detected by the negativity criterion~\cite{Horo97}. These PPT states were detected using the so-called Range criterion. It states that if $\rho$ is separable then there exist a separable vector $\ket{\mathbf{e}f}$ such that $\ket{\mathbf{e}f} \in R(\rho)$ and $\ket{\mathbf{e}^\star f} \in R(\rho^{T_A})$, with $\ket{\mathbf{e}^\star f}$ indicating complex conjugation in the first subsystem and $R$ is used to indicate the range of an operator. Another way of expressing this criterion is to say that if there is no separable vector $\ket{\mathbf{e}f}$ such that:
\begin{eqnarray}\label{rangecrit}
\braket{k_i|\mathbf{e}f}=0 \ \forall \ket{k_i}\in K(\rho) \quad and \quad \braket{\kappa_i|\mathbf{e}^\star f}=0 \ \forall \ket{\kappa_i}\in K(\rho^\Gamma),
\end{eqnarray}
with $K$ representing the kernel of an operator, then $\rho$ must be entangled. Bound entangled states satisfying (\ref{rangecrit}) are called edge states \cite{Lewenstein00}.

\section{Bound entanglement in the JC model}\label{sec:bound}
In this section we show that there are bound entangled states that satisfy (\ref{comm}) and thus from now on, we only consider states that are PPT. We will assume that the local unitary operation in equation (\ref{local}) has been applied and thus $c_n=c_n^\star  \geq 0$. 
To reduce the complexity of the problem and facilitate the proof it is convenient to apply the following local filtering operation to the state:
\begin{eqnarray}
F &=&  \mathbb{I} \otimes F_B\\
F_B&=&\sum_{n=0}^{N-1} \frac{1}{\sqrt{b_n}} \ket{n}\bra{n}. \nonumber
\end{eqnarray}
Notice that if there is some $b_m=0$ then from (\ref{pos}) and (\ref{postrans}) $c_{m+1}=c_m=0$. In such case the density matrix can be split as follows:
\begin{eqnarray}
\rho&=&\rho_{[0:m-1]}+\rho_{[m]}+\rho_{[m+1:N-1]}\\
\rho_{[0:m-1]}&=& \sum_{n=0}^{m-1} \left( a_n \ket{\g n}\bra{\g n}+b_n \ket{\x n}\bra{\x n} \right)\\
&&+\sum_{n=1}^{m-1} c_n \left(\ket{\x n-1}\bra{\g n}+\ket{\g n}\bra{\x n-1}\right) \nonumber \\
\rho_{[m]}&=& a_m \ket{\g m} \bra{\g m}\\
\rho_{[m+1:N-1]}&=& \sum_{n=m+1}^{N-1} \left( a_n \ket{\g n}\bra{\g n}+b_n \ket{\x n}\bra{\x n} \right)\\
&&+\sum_{n=m+2}^{N-1} c_n \left(\ket{\x n-1}\bra{\g n}+\ket{\g n}\bra{\x n-1}\right) \nonumber
\end{eqnarray}
where $\rho_{[0;m-1]}$ is a state supported in $2 \times m-1$, $\rho_{[m+1,N-1]}$ is a state supported in $2\times (N-m)$ and $\rho_{[m]}$ is a separable state. Because of this the problem is reduced to 2 lower rank density matrices with the same structure as the original one plus a trivially separable state. Now, without loss of generality we assume that all the populations of the density matrix are nonzero and after applying the above filtering operation to $\rho$ one obtains:
\begin{eqnarray}\label{filter}
\rho \rightarrow \sigma = F \rho F^\dagger &=& \sum_{n=0}^{N-1}\left( x_n^2 \ket{\g n}\bra{\g n}+ \ket{\x n}\bra{\x n}\right)\\
&&+ \sum_{n=1}^{N-1} y_n \left(\ket{\x n-1}\bra{\g n}+\ket{\g n}\bra{\x n-1}\right)\nonumber 
\end{eqnarray}
\begin{eqnarray}
x_n^2&=&a_n/b_n
\end{eqnarray}
\begin{eqnarray}\label{yn}
y_n&=&c_n/\sqrt{b_{n-1} b_n}.
\end{eqnarray}
Local filtering operations map entangled states to entangled states and separable states to separable states and thus $\sigma$ is still positive and is PPT. The positivity and PPT conditions equivalent to equations (\ref{pos},\ref{postrans}) are now 
\begin{eqnarray}\label{ppt}
y_i \leq \min(x_i,x_{i-1})
\end{eqnarray}
or equivalently\footnote{
We note that there is a slight abuse of notation in writing (\ref{ppts}) since there are only $N-1$ variables $y_i$ with the index $1\leq i \leq N-1$ (see (\ref{yn}) in which the $y_i$ are defined in terms of the $c_i$). Because of this when $i=0$ or $i=N-1$ in (\ref{ppts}) the equation becomes meaningless since $y_0$ or $y_N$ are not defined (they are not part of the problem) and indeed going back to equation (\ref{ppt}), $x_0$ and $x_{N-1}$ only need to satisfy $x_0 \geq y_1$ and $x_{N-1}\geq y_{N-1}$. To solve this inconvenience one could simply define $\max(y_0,y_1)\equiv y_1$ and $\max(y_{N-1},y_N)\equiv y_{N-1}$ or introduce two fictitious or auxiliary (in the sense that they are not part of the problem) quantities $y_0$ and $y_N$ with the values $y_0 \equiv y_1$ and $y_N \equiv y_{N-1}$.
}:
\begin{eqnarray}\label{ppts}
x_i \geq \max(y_i,y_{i+1}).
\end{eqnarray}
Further, notice that the above matrix can be split as a trivially separable part $\sigma_s$ plus a  PPT state $\tau$ that satisfies $r(\tau)+r(\tau^\Gamma) \leq 3N$ with $r(\omega)$ being the dimensionality of the range of $\omega$:
\begin{eqnarray}\label{pptness}
\sigma&=&\sigma_s+\tau\\
\sigma_s&=&\sum_{n=0}^{N-1}\left( x_n^2-\max(y_n^2,y_{n+1}^2) \right) \ket{\g n}\bra{\g n} \\
\tau &=&\sum_{n=0}^{N-1}\left(\max(y_n^2,y_{n+1}^2)\ket{\g n}\bra{\g n} +\ket{\x n}\bra{\x n}\right)\\
&&+ \sum_{n=1}^{N-1} y_n \left(\ket{\x n-1}\bra{\g n}+\ket{\g n}\bra{\x n-1}\right).\nonumber
\end{eqnarray}
From the above it is clearly seen that the separability properties of $\sigma$ are the same as those of $\tau$.\\
Having transformed the original states (\ref{state}) to the form (\ref{pptness}) in what follows we will show that there are bound entangled states that satisfy (\ref{comm}). Since we are only interested in showing the existence of bound entangled states of the type (\ref{state}) we will consider the smallest dimension in which this is possible, \emph{i.e.}, $N=4$. The study of the entanglement properties of these states for arbitrary $N$ will be presented elsewhere \cite{NicoAnna13}, nevertheless we mention that based on the same ideas that will be presented in the following paragraphs it is possible to show that there are bound entangled states for arbitrary $N$. For the sake of concreteness we will only examine the cases in which: 
\begin{eqnarray}\label{ineq}
y_{i}\leq y_{i+1},
\end{eqnarray}
thus,
\begin{eqnarray}\label{N4}
\tau=\left(
\begin{array}{cccc|cccc}
 y_1^2 &  &  &  &  &  &  &  \\
  & y_2^2 &  &  & y_1 &  &  &  \\
  &  & y_3^2 &  &  & y_2 &  &  \\
  &  &  & y_3^2 &  &  & y_3 &  \\
\hline
  & y_1 &  &  & 1 &  &  &  \\
  &  & y_2 &  &  & 1 &  &  \\
  &  &  & y_3 &  &  & 1 &  \\
  &  &  &  &  &  &  & 1 \\
\end{array}
\right)
\end{eqnarray}
The dimensions of the ranges of $\tau$,  and $\tau^\Gamma$,  are given by $r(\tau)=7$ and $r(\tau^\Gamma)=5$ if none of the inequalities (\ref{ineq}) are saturated. It is known \cite{kraus00} that the matrix is separable if $r(\tau)=4$ or $r(\tau^\Gamma)=4$, therefore, one should proof if there are enough product vectors in the range so that the rank of the matrices diminishes appropriately. The $3$ vectors in the Kernel of $\tau^\Gamma$, are given by:
\begin{eqnarray}\label{consn}
\ket{\phi_1}&=&-\ket{\g 0}+y_1\ket{\x 1},\\
\ket{\phi_2}&=&-\ket{\g 1}+y_2\ket{\x 2},\nonumber\\
\ket{\phi_3}&=&-\ket{\g 2}+y_3\ket{\x 3}\nonumber,
\end{eqnarray}
whereas the vector in the kernel of $\tau$, is simply:
\begin{eqnarray}\label{cons1}
\ket{\chi_3}&=&-\ket{\g  3}+y_3\ket{\x 2}.
\end{eqnarray}
Now we need to find a separable vector $\ket{\mathbf{e}f}$ such that 
\begin{eqnarray}
\braket{\chi_3|\mathbf{e}f}=0  \quad and \quad \braket{\phi_n|\mathbf{e}^\star  f}=0 \ \forall n. 
\end{eqnarray}
It is found that the unique (up to a phase and normalization) separable vector that is orthogonal to (\ref{consn}) and that upon complex conjugation in the qubit system is orthogonal to (\ref{cons1}) is:
\begin{eqnarray}\label{states}
\ket{\mathbf{e}f}_{\theta}&=&\left(\ket{\g}+\frac{e^{i\theta}}{y_3}\ket{\x} \right)\nonumber\\
&&\otimes \left(\frac{y_1 y_2}{y_3} \ket{0}+y_2 e^{i \theta} \ket{1}+y_3 e^{i 2\theta}\ket{2}+ e^{i 3\theta} y_3\ket{3}\right).
\end{eqnarray}
Now, to show that there bound entangled states of the type defined by (\ref{state}) we will look at a subset of the states defined by (\ref{N4}). Up to now we assumed that none of the inequalities (\ref{ineq}) was saturated. To construct our bound entangled state we will look at the case when one of them is saturated, to be precise, we fix $y_2=y_1$ and thus our state is simply:
\begin{eqnarray}\label{bound4}
\tau(y_1,y_2)=\left(
\begin{array}{cccc|cccc}
 y_2^2 &  &  &  &  &  &  &  \\
  & y_2^2 &  &  & y_2 &  &  &  \\
  &  & y_3^2 &  &  & y_2 &  &  \\
  &  &  & y_3^2 &  &  & y_3 &  \\
\hline
  & y_2 &  &  & 1 &  &  &  \\
  &  & y_2 &  &  & 1 &  &  \\
  &  &  & y_3 &  &  & 1 &  \\
  &  &  &  &  &  &  & 1 \\
\end{array}
\right).
\end{eqnarray}
With this new constraint the dimension of the range of $\tau$ is decreased by one and a new vector appears in the Kernel of $\tau$:
\begin{eqnarray}\label{ketn}
\ket{\chi_{1}}=-\ket{\g 1}+y_{1}\ket{\x 0}.
\end{eqnarray}
As we mentioned before, the vector (\ref{states}) is the \emph{only} separable vector that is orthogonal to (\ref{consn}) and (\ref{cons1}) and such constraints are not modified by assuming $y_2=y_1$. Nevertheless this last assumption also implies that vector (\ref{states}) must also be orthogonal to (\ref{ketn}) for the state $\tau$ to be separable. The inner product between the separable vector (\ref{states}), and (\ref{ketn}) is:
\begin{eqnarray}
\braket{\chi_{1}|\mathbf{e} f}_{\theta}=e^{i \theta} y_2\left(\frac{y_2}{y_3}-1 \right) \neq 0.
\end{eqnarray}
Since we are free to take $y_2 < y_3$ we conclude that in such case there is no separable vector $\ket{\mathbf{e}f}$ that satisfies the hypothesis of the range criterion (\ref{rangecrit}) and hence we conclude that \emph{there exist bound entangled states of the type defined by (\ref{comm}) for $N > 3$}.  \\
We point out that the above argument requires at least three $y_i$ and thus not surprisingly will only work for $N > 3$. In \ref{app:hull} we construct the convex hulls for the PPT separable states for $N=2,3$. In \ref{app:crit} we explicitly evaluate two often used criteria that have been shown to detect some bound entangled states and numerically show that they are unable to detect bound entanglement for the states considered here.

\section{Generation of bound entanglement using the JC interaction}\label{sec:dyn}
In the last section it was shown that there are bound entangled states compatible with the JC symmetries. In this section we shall show that the JC interaction can be used to generate bound entanglement from uncorrelated states. We will study the interaction of an atom prepared in the unpolarized state:
\begin{eqnarray}
 \rho(0)_A = (1-\lambda) \ket{\g} \bra{\g}+\lambda \ket{\x}\bra{\x}
\end{eqnarray}
where $\lambda$ represents the probability of having the atom in the excited state and a field prepared in the thermal state:
\begin{eqnarray}
 \rho(0)_B=\sum_{n=0}^{\infty} p_n \ket{n}\bra{n}
\end{eqnarray}
with 
\begin{eqnarray}
p_n=\frac{m^n}{(1+m)^{n+1}}
\end{eqnarray}
and $m=\braket{ a ^\dagger  a}$ is the mean number of photons in the field. The atom-cavity system starts in the product state:
\begin{eqnarray}
 \rho(0)= \rho(0)_A \otimes \rho(0)_F
\end{eqnarray}
The dynamics of this type of states under the resonant JC hamiltonian ($\Delta=0$ in equation (\ref{HJC})) has been studied by Scheel \emph{et.al.} in \cite{Plenio03}. They find that in the $(\lambda,m)$ parameter space there are three separate regions:
\begin{enumerate}
\item In the first region the state becomes free entangled immediately after the interaction between the atom and the field starts, \emph{i.e.}, for $t>0$ the state does not have a positive partial transpose.
\item In the second region the state becomes free entangled only after some finite $\bar t$, \emph{i.e.} for some finite $\bar t$ the state has positive partial transpose and then after this time the state becomes free entangled.
\item In the third region they find that the state is PPT for all times.
\end{enumerate}
In this section we will be interested in the cases for which the state remains PPT for a finite time after the interaction starts, \emph{i.e.} regions (ii) and (iii).\\
To understand how bound entanglement is generated we first write explicitly the time evolution generated by $H_{JC}$ in the \emph{resonant} case:
\begin{eqnarray}
\rho(t)&=&\exp\left(-i H_{JC} t \right) \rho(0) \exp\left(i H_{JC} t \right)\\&=&
\alpha_0^- \ket{ \g 0}\bra{ \g 0}+\sum_{n=1}^\infty f_n \left(\alpha_{n}^- \pr{\g n}+\alpha_{n}^+ \pr{\x n-1} \right.\nonumber\\
&&\quad \quad \quad \quad \quad \quad \quad \left. +\beta_n \left[ \ket{\g n}\bra{\x n-1}+ \ket{\x n-1}\bra{\g n} \right] \right) \nonumber
\end{eqnarray}
with:
\begin{eqnarray}
f_n&=&\frac{1}{2} m^{n-1} (m+1)^{-n-1}\\
\alpha_{n}^{\pm}&=&m+\lambda \pm (\lambda+m (2 \lambda-1)) \cos \left(2 g \sqrt{n} t\right)\\
\beta_n&=&(m (2 \lambda-1)+\lambda) \sin \left(2 g \sqrt{n} t\right).
\end{eqnarray} 
Remembering the notation introduced in equations (\ref{not1}) and (\ref{not2}) it is found that the nonzero elements of $\rho$ are:
\begin{eqnarray}
a_n=f_n \alpha_{n}^-, \quad b_n=f_{n+1} \alpha_{n+1}^+, \quad c_n=|f_n \beta_n|,
\end{eqnarray}
and from them the $y_n$ defined in equation (\ref{yn}) can be found. We will be interested in finite times short enough that the Taylor expansion
\begin{eqnarray}\label{yshort}
y_n&=&\frac{f_n \beta_n}{\sqrt{f_n \alpha_{n}^+ f_{n+1} \alpha_{n+1}^+}} \\
&\approx& \left|\sqrt{n}T +\mathcal{O}\left(T^3\right)  \right|
\end{eqnarray}
with
\begin{eqnarray}
T &=& \left|\frac{\lambda+ m (2 \lambda-1)}{\lambda \sqrt{m (m+1)} } \right| g t
\end{eqnarray}
remains valid. Finally, since we are only interested in the existence of bound entanglement we will truncate the photon ladder in 3 photons thus including only the lowest 4 Fock states. This corresponds to a local operation in the photonic system and thus cannot create entanglement. In particular that means that if the truncated density operator is entangled then it can be concluded that the full state is entangled.\\
Applying the truncation and decomposing the state according to equations (\ref{filter}) and (\ref{pptness}) it is found that the bound entanglement properties of the state $\rho(t)$ are equivalent to those of the state (\ref{N4}) with the $y_i$ given in equation (\ref{yshort}). Now to show that such state is entangled we note that it can be written as:
\begin{eqnarray}
\tau=\tau_1+\lambda \ket{\mathbf{e}f}\bra{\mathbf{e}f}_{\theta=0}
\end{eqnarray}
with $\ket{\mathbf{e}f}_{\theta}$ given by equation (\ref{bound4}) and $\lambda=\frac{27}{43} T^2$ and $\tau_1$ being a PPT state that satisfies:
\begin{eqnarray}
K(\tau_1)&=&span\left(\ket{\chi_3},\ket{\zeta} \right)\\
K(\tau_1^\Gamma)&=&span\left(\ket{\phi_1}, \ket{\phi_2},\ket{\phi_3} \right)
\end{eqnarray}
with $\ket{\chi_3}$ and the $\ket{\phi_i}$ given in equations (\ref{cons1}) and (\ref{consn}) and 
\begin{eqnarray}
\ket{\zeta}&=&\frac{\sqrt{6}}{ T } \ket{\g 0}+ \frac{2 \sqrt{2}}{ T } \ket{\g 1}+\frac{\sqrt{3}}{ T}\ket{\g 2}+\frac{\sqrt{3}}{ T }\ket{\g 3} -\sqrt{2}\ket{\x 0}+3 \ket{\x 3}.
\end{eqnarray}
To show that $\tau_1$ is entangled it is sufficient to notice that the only separable vector that satisfies being orthogonal to $\ket{\chi_3}$ and that upon conjugation on the first subsystem is orthogonal to $\ket{\phi_1},\ket{\phi_2},\ket{\phi_3}$ is $\ket{\mathbf{e}f}_{\theta}$. Nevertheless, it is easily seen that:
\begin{eqnarray}
\braket{\zeta|\mathbf{e}f}_\theta \neq 0 \quad \forall \theta
\end{eqnarray}
and thus we conclude that $\tau_1$ is an edge entangled state and that $\tau$ is bound entangled since it is a mixture of an entangled state and a pure separable one.\\
This result implies that for any value of the parameters $m$ and $\lambda$ (other than $m=0$ or $\lambda=0$ in which case the Taylor expansion (\ref{yshort}) is meaningless) the state always becomes entangled for short times after the JC interaction between the atom and field starts. If the initial state happens to be sitting in the region (i) mentioned at the beginning of this section it will become free entangled. More interestingly for our purposes, \emph{if it happens to be sitting in region (ii) or (iii) it will become bound entangled}.

\section{Conclusion}\label{sec:conc}
In this paper we have analyzed the entanglement properties in the JC model. To this aim we have first used the conservation of the number operator to formalize the mixed states that naturally occur when  a two-level atom interacts with an electromagnetic field in a cavity evolving under the JC dynamics. We have first examined the limitations of the entanglement criteria used so far to study these states. Then, to the best of our knowledge, we have for the first time demonstrated (analytically) that bound entanglement exist in such model and the failure of some of the criteria used so far to detect such type of entanglement. Finally, we shown that the JC interaction can be used to generate bound entangled states starting from mixed uncorrelated ones. Our results have implications for all systems whose dynamics can be approximated by a Jaynes-Cummings Hamiltonian, such as ion traps and cavity/circuit QED.

\section*{Acknowledgments}\label{sec:ack}
We are indebted to M. Lewenstein for crucial enlightening discussions. N.Q. gratefully acknowledges valuable discussions with O. Gittsovich, G. T\'oth, D. \v{Z}. Djokovi{\'c}, E. Wolfe, D.F.V. James and financial support from the National Sciences and Engineering Research Council of Canada. A.S. acknowledges support from the Spanish MICINN FIS2008-01236  European Regional Development Fund (FEDER), discussions with R. Quesada and specially the kind hospitality of J. H. Thywissen and the Univ. of Toronto where this work was initiated.

\appendix
\section{Bound entanglement detection criteria}\label{app:crit}
In the entanglement detection criteria to be evaluated in this appendix it is assumed that the states are normalized:
\begin{eqnarray}
\tr(\rho)=\sum_{n=0}^{N-1}\left(a_n+b_n \right)=1.
\end{eqnarray}
it will also be convenient to write the marginals (reduced density matrices) of the qubit and the qudit:
\begin{eqnarray}\label{marginals}
\rho_{A}=\tr_{B}\left(\rho\right)=\sum_{\ii = \g}^{\x} \alpha_{\ii} \ket{\ii}\bra{\ii}, \quad 
\rho_{B}=\tr_{A}\left(\rho\right)=\sum_{n=0}^{N-1} \beta_n \ket{n}\bra{n},
\end{eqnarray}
where the coefficients are given by:
\begin{eqnarray}\label{coeff}
\alpha_{\g}=\sum_{n=0}^{N-1} a_n, \quad \alpha_{\x}=\sum_{n=0}^{N-1} b_n, \quad 
\beta_n=a_n+b_n.
\end{eqnarray}
They satisfy the normalization $\alpha_\x +\alpha_\g =\tr\left(\rho \right)=\sum_{n=0}^{N-1} \beta_n$.\\
The first criterion that we evaluate is the computable cross-norm or realignment (CCNR) criterion. The CCNR criterion states that if $\rho$ is separable, then the following inequality must hold \cite{Rudolph03,Guhne09},
\begin{equation}\label{CCN}
    \|\mathcal{R}(\rho)\|\leq1,
\end{equation}
here $\|\cdot\|$ stands for the trace norm (\emph{i.e.} the sum of the singular values). The realignment operation is defined as $\mathcal{R}(A\otimes B)=|A\rangle\langle B^{*}|$, with scalar product $\langle B|A\rangle=\mathrm{tr}(B^{\dag} A)$ in the Hilbert Schmidt space of operators. For the states considered here the singular values are:
\begin{eqnarray}
s(\mathcal{R}(\rho))=\{ |\vec c|, |\vec c|, x_+,x_-\},
\end{eqnarray}
with
\begin{eqnarray*}
x_{\pm}=\sqrt{\frac{|\vec a|^2+|\vec b|^2\pm\sqrt{(|\vec a|^2-|\vec b|^2)^2+4 (\vec a \cdot \vec b)^2}}{2}},
\end{eqnarray*}
and where the following real vectors are defined in terms of the populations and coherences of the density matrix:
\begin{eqnarray}
\vec a& =& \left\{a_0,\ldots, a_{N-1}\right\} \in \mathbb{R}^N\\
\vec b& =& \left\{b_0,\ldots, b_{N-1}\right\} \in \mathbb{R}^N \nonumber\\
\vec c& =& \left\{|c_1|,\ldots,|c_{N-1}|\right\} \in \mathbb{R}^{N-1}. \nonumber
\end{eqnarray}
and $|\vec x|=\sqrt{\vec x \cdot \vec x}$.
Note that the four singular values will be nonzero unless $|\vec c|=0$ \emph{i.e.} $c_n=0 \ \forall n$ or $(\vec a \cdot \vec b)^2=|\vec a|^2 |\vec b|^2$ \emph{i.e.} $a_n= w b_n \ \forall n$.
If $\|\mathcal{R}(\rho)\|>1$ the state $\rho$ will be entangled, explicitly one finds that:
\begin{eqnarray}
\|\mathcal{R}(\rho)\|&=&2 |\vec c|+x_+ + x_-=2|\vec c|+\sqrt{|\vec a|^2+|\vec b|^2+2\sqrt{|\vec a|^2 |\vec b|^2-(\vec a \cdot \vec b)^2}}.
\end{eqnarray}
In \cite{Kai05} it is shown that $ \|\mathcal{R}(\rho)\|-1$ will provide a lower bound of the concurrence that is \emph{in principle} independent of whether the state has positive PT or not. We generated more than ten million states for $N=4$ of which a million and a half where PPT. In all cases $\mathcal{N}(\rho)\geq \max\{ \| \mathcal{R}(\rho) \| -1,0 \}$.
Finally, one can improve the entanglement detection capabilities of the CCNR criterion by the following corollary of the CM method \cite{Zhang08,Oleg08}. The corollary states that if $\rho$ is separable then:
\begin{eqnarray}\label{zzzg}
\|\mathcal{R}(\rho-\rho_A \otimes \rho_B)\|^2 
\leq \left(1-\tr\left\{\rho_A^2 \right\} \right) \left(1-\tr\left\{\rho_B^2 \right\} \right).
\end{eqnarray}
For the states considered here one can show that:
\begin{eqnarray}
\|\mathcal{R}(\rho-\rho_A \otimes \rho_B)\|=2|\vec c|+\sqrt{2}|\alpha_{\x} \vec a-\alpha_{\g} \vec b|.
\end{eqnarray}
The coefficients $\alpha_{\ii}$ are given in equations (\ref{coeff}). To test if this criterion could detect any bound entanglement we generated a million and a half PPT states in $2 \times 4$ and none of them violated inequality (\ref{zzzg}). To confirm that indeed the CM corollary (which is always stronger than the CCNR criterion \cite{Oleg08}) does not detect the states (\ref{bound4}) we considered $N=4$ and looked at the two parameter family of normalized states  $\rho(y_2,y_3) = \tau(y_2,y_3)/\tr(\tau(y_2,y_3))$ in the range $0 < y_2 <y_3 \leq 10$ and in none of the parameter values the inequality (\ref{zzzg}) was violated.

\section{Convex hulls for $N=2,3$}\label{app:hull}
As it is well known in these cases being PPT \emph{implies} separability. In this appendix we explicitly construct the convex hull of the PPT states.
For future convenience we will introduce the following projector CP-map that acts on states $\omega$:
\begin{eqnarray}
\mathcal{P}(\omega)\equiv \frac{1}{2N}\sum_{k=0}^{2N-1} \exp\left(i \frac{\pi k  \Pi}{N}\right)\omega \exp\left(-i \frac{\pi k \Pi}{N}\right).
\end{eqnarray}
The superoperator $\mathcal{P}$ has the property that for any $\omega$:
\begin{eqnarray}
\left[\mathcal{P}(\omega), \Pi \right]=0.
\end{eqnarray}
Thus $\mathcal{P}$ projects onto the subspace defined by equation (\ref{comm}), it satisfies $\mathcal{P}( \mathcal{P}(\omega))=\mathcal{P}(\omega)$, cannot generate entanglement since it can be implemented with local operations and classical communication and it maps density operators to density operators.\\
In the $N=2$ case one has:
\begin{eqnarray}\label{22}
\tau=
\left(
\begin{array}{cc|cc}
 y_1^2 &  &  &  \\
  & y_1^2 & y_1 &  \\
\hline
  & y_1 & 1 &  \\
  &  &  & 1 \\
\end{array}
\right).
\end{eqnarray}
It is easy to see that there are separable product in vectors $\ket{\mathbf{e} f}$ in the range of $\tau$, $R(\tau)$ such that $\ket{\mathbf{e}^\star f}$ is in the range of $\tau^\Gamma$, $R(\tau^\Gamma)$. One of such product vectors is:
\begin{eqnarray}
\ket{\mathbf{g} h}=( \ket{\g}+\frac{1}{y_1}\ket{\x})\otimes(y_1 \ket{0}+y_1\ket{1}).
\end{eqnarray}
It is easily shown that $\mathcal{P}(\ket{\mathbf{g}h}\bra{\mathbf{g}h})$ is precisely $\tau$ thus showing the convex hull of the separable state. Not surprisingly the four vectors that are added to construct $\mathcal{P}(\ket{\mathbf{g}h}\bra{\mathbf{g}h})$ and obtain (\ref{22}) are the same four vectors that one would obtain by using the Wootters formula for a system of 2 qubits\cite{Wootters98}.\\
For $N=3$ one has two possibilities for $\tau$:
\begin{eqnarray}
\tau= \left(
\begin{array}{ccc|ccc}
 y_1^2 &  &  &  &  &  \\
 & y_1^2 &  & y_1 &  &  \\
  &  & y_2^2 &  & y_2 &  \\
\hline
  & y_1 &  & 1 &  &  \\
  &  & y_2 &  & 1 &  \\
  &  &  &  &  & 1 \\
\end{array}
\right) \quad or \quad 
\left(
\begin{array}{ccc|ccc}
 y_1^2 &  &  &  &  &  \\
 & y_2^2 &  & y_1 &  &  \\
  &  & y_2^2 &  & y_2 &  \\
\hline
  & y_1 &  & 1 &  &  \\
  &  & y_2 &  & 1 &  \\
  &  &  &  &  & 1 \\
\end{array}
\right)
\end{eqnarray}
depending on whether $y_1 \geq y_2$ or $y_2 \geq y_1$. In the first case a vector that is in $R(\tau)$ and $R(\tau^\Gamma)$ is simply:
\begin{eqnarray}\label{33}
\ket{\mathbf{g}h}= \left( \ket{\g}+\frac{1}{y_1}\ket{\x}\right)\otimes(y_1\ket{0}+y_1\ket{1}+y_2\ket{2})
\end{eqnarray}
upon subtracting $\mathcal{P}(\ket{\mathbf{g}h}\bra{\mathbf{g}h})$ one obtains $\left(1-\frac{y_2^2}{y_1^2} \right)\ket{\x}\bra{\x}\otimes \ket{2}\bra{2}$. In the case where $y_2 \geq y_1$  one finds a product vector in $R(\tau)$ and $R(\tau^\Gamma)$ by swapping $y_1$ and $y_2$ in (\ref{33})upon subtracting $\mathcal{P}(\ket{\mathbf{g}h}\bra{\mathbf{g}h})$ one obtains $\left(1-\frac{y_1^2}{y_2^2} \right)\ket{\g}\bra{\g}\otimes \ket{2}\bra{2}$ thus completing the construction for $N=2,3$. We note that the technique presented here using the projector $\mathcal{P}$ is equivalent to the uniform mixing/averaging over all phases used to find separability proofs for symmetric mixed states of $N$ qubits\cite{elie13}.

\section*{References}
\bibliographystyle{iopart-num}

\bibliography{main}

\end{document}